\documentclass[letterpaper,12pt]{article}
\usepackage{amsmath}
\usepackage{amsfonts}
\usepackage{amssymb}
\usepackage{graphicx}
\usepackage{physics}
\usepackage{latexsym}
\usepackage{epsfig}
\usepackage{pstricks}
\usepackage{stmaryrd}
\usepackage{rotating}
\usepackage[english]{babel}
\usepackage{setspace}
\usepackage[utf8]{inputenc}
\usepackage{natbib}
\usepackage[T1]{fontenc}
\usepackage{lmodern}
\usepackage{enumitem}
\usepackage{csquotes}
\usepackage{amsthm}
\usepackage{xcolor}
\usepackage[margin=1in]{geometry}
\begin{document}
\begin{center}	
\begin{Large}
\textbf{On the detection of absolute velocity in a Newtonian universe}\\
\end{Large}
\end{center}

\begin{center}
\begin{large}
Jorge Manero, Ricardo Muciño and Elias Okon\\
\end{large}
\textit{Universidad Nacional Aut\'onoma de M\'exico, Mexico City, Mexico.}\\[1cm]
\end{center}

As a fundamental arena for the development of his dynamics, Newton postulated the existence of \emph{absolute space}, in which bodies innately possess \emph{absolute velocity}. Despite this, Newton argued that, although real, absolute properties cannot be detected. Since then, the claim that absolute velocity would be undetectable in such a Newtonian universe has been generally accepted. Here, we show that standard arguments for such a claim, beginning with the one offered by Newton himself, beg the question. We conclude that there are no formal reasons to believe that absolute velocity would be undetectable in a Newtonian universe.\\


\begin{flushright}
\textit{It is the theory which decides what can be observed.}\\ ---Albert Einstein\footnote{As quoted by W. Heisenberg in \cite{heisenberg1971physics}.} 
\end{flushright}

\onehalfspacing
\section{Introduction}

As an underlying stage for the formulation of classical mechanics, Newton postulated the existence of \emph{absolute space}---a fixed, immutable backdrop against which all motion occurs. Within such a framework, physical bodies invariably possess \emph{absolute velocity}, independent of any relative comparisons with other objects; velocity is not merely relational but has an objective character when considered in reference to absolute space itself.\footnote{Contemporary presentations of Newtonian mechanics eschew absolute space in favor of so-called Neo-Newtonian or Galilean spacetime, in which all velocities are relative.} 

In spite of the fact that, in this framework, absolute space and absolute velocity are real, objective features, Newton argued that they cannot be directly observed or measured through empirical means. This position led to the widely accepted conclusion that absolute velocities are undetectable within a Newtonian universe, a claim that has been generally endorsed by both Newton's contemporaries and subsequent generations of physicists and philosophers.  

Here, we show that standard arguments for this claim based on the purely formal elements of the framework, given by Newton's laws and the structure of absolute space, ultimately rely on circular arguments. In other words, these arguments (inadvertently) \emph{assume} the undetectability of absolute properties, rather than demonstrating it as a necessary consequence of the purely formal apparatus of Newtonian mechanics. Consequently, we conclude that there is no compelling reason to accept the claim that absolute velocity is inherently undetectable within the Newtonian framework.

More concretely, what we show is that, within a Newtonian universe, absolute and relative velocities exhibit the same relevant formal features regarding detection: both allow for adequate measurement protocols, the establishment of necessary correlations, and the satisfaction of all required counterfactuals. Thus, there is no way, based solely on the formal aspects of Newtonian theory, to claim that absolute velocities are undetectable; any insistence that relative quantities possess some extra ingredient would amount to an additional stipulation without formal justification. Our point is not that absolute velocity is detectable, but that standard arguments purporting to prove its undetectability from the formal framework of Newtonian theory alone are deficient. This stands in contrast to the standard view that Newtonian dynamics itself guarantees the impossibility of detecting absolute velocity. While absolute and relative quantities certainly differ in many respects, none of these differences supports the usual claim that absolute velocity must be empirically inaccessible.

As Galileo convincingly argued---and many experiments since have attested---it is true that, in our actual universe, we don't seem able to detect absolute velocity. However, it is very important not to confuse such an \emph{empirical} fact, with the purely \emph{theoretical} question of whether absolute velocity can be shown, from purely formal aspects of Newtonian mechanics, to be undetectable in a \emph{hypothetical} Newtonian universe. Therefore, in order to assess the theoretical question at hand, one should be very careful not to employ any actual empirical data, as this pertains to our universe, not the hypothetical one under consideration. That is, to address the issue, one must keep within the bounds of the theoretical framework surveyed. The fact that our actual universe is patently not Newtonian might, in turn, seem to greatly diminish the practical usefulness of our analysis. However, as we argue below, the apparently purely academic exercise of exploring whether absolute velocity could be shown to be undetectable in a Newtonian universe yields important lessons for real world issues, both in physics and general philosophy of science.

The issue of the detectability of absolute velocity has recently attracted some attention. In particular, in \cite{murgueitio21}, the influential defense in \cite{Roberts} of the claim that absolute velocity is not measurable is challenged with the proposal for an alternative analysis of measurement, leading to the conclusion that absolute velocity is measurable in at least one world. The conclusions of our work are stronger, as they do not depend on a potentially controversial analysis of measurement, and they apply to all worlds. The claims in \cite{murgueitio21} are challenged in \cite{Jacobs}, with a further discussion in \cite{Luc} and \cite{murgueitio24}. \cite{Dasgupta2016} and \cite{wallace}, on the other hand, explore the more general question of the observability of variant properties. While both of these works seem to agree with us on a number of issues, they ultimately draw different conclusions. \cite{wallace}, in particular, seems to agree with most of our arguments regarding the analysis in \cite{Roberts} but, by imposing what we take to be an unjustified restriction on possible measurement records, ultimately arrives at the opposite of our conclusion.

Our manuscript is organized as follows. In section \ref{ASA}, we assess the standard arguments in favor of the claim that absolute velocity would be undetectable in a Newtonian universe and, in section \ref{Det}, we make some general remarks about the notion of detection. Next,  in section \ref{Rob}, we assess the influential analyses of the issue presented in \cite{Roberts} and \cite{wallace} and, in section \ref{Mis}, we clarify some potential lingering misunderstandings. Finally, in section \ref{Cons} we examine some broader consequences of our analysis and, in section \ref{Conc}, we offer our conclusions.

\section{Assessing the standard arguments}
\label{ASA}

Consider a Newtonian universe, where space is absolute and Newton's laws of motion are valid. It is commonly accepted that absolute velocity is undetectable in such a framework, as only relative motion is thought to have observable consequences. In this section, we examine standard arguments for this claim arising from the purely formal elements of the framework, given by Newton's laws and the structure of absolute space---beginning with Newton's own---and show that they rely on assumptions that presuppose their conclusions. In doing so, we conclude that the claim of absolute velocity's undetectability does not follow from the formal structure of the Newtonian universe, but instead rests on question-begging reasoning.
	
The gist of the standard arguments is the following. One first shows that, if all forces are either contact forces or forces that only depend on relative positions and velocities of bodies, then these relative quantities are not affected by a Galilean transformation. As Newton put it for the case of Galilean boosts under the influence of forces of the sort described above, ``the motions of bodies included in a given space are the same among themselves, whether that space is at rest, or moves uniformly forwards in a right line without any circular motion'' \cite[pp. 20-21]{Newton}. From the fact that all relative quantities are Galilean-invariant, it is concluded that all absolute properties---among them, velocities---are undetectable.
	
However, these arguments are circular. Before discussing why, we note that the restriction to only consider contact forces, or forces that only depend on relative quantities, is somewhat \emph{artificial} in a Newtonian universe. In such a universe, absolute properties are real and objective, so there is no \emph{a priori} reason to impose that such properties cannot influence forces. And, as pointed out in \cite[Sec. 3.2] {Brown}, without such a restriction, absolute velocities could be detected.\footnote{It could be thought that allowing for absolute properties to affect forces would necessarily break either Newton's third law or the assumption of homogeneity and isotropy of absolute space. However, that is not the case. Consider, for instance, a Newtonian universe populated by point particles, with a pairwise force such that the coupling constant of the force is a function of the magnitude of the absolute velocity of the center of mass of the pair of particles involved. For concreteness, consider an attractive, inverse-square, mass-proportional force, such as Newtonian gravity, but in which $g$, the analog of Newton's $G$, is a function of the magnitude of the absolute velocity of the center of mass of the pair of particles affected. That is, where the magnitude of the force between particles $i$ and $j$ is given by
\begin{equation}
F_{ij}=g(|v^{cm}_{ij}|)\frac{m_i m_j}{r_{ij}^2} .
\end{equation}
It seems clear that such a force, neither disturbs Newton's third law, as for any pair of particles it is the case that the forces exerted between them are opposite and equal in magnitude, nor that it severs the homogeneity and isotropy of absolute space, as all points and directions are treated equally.}

At any rate, the point we want to make is that, even if, by stipulation, all forces only depend on relative quantities, it still does not follow that absolute velocities are undetectable. This is because, as mentioned above, all one can show by assuming forces to depend on relative quantities, is that all relative quantities are Galilean invariant. However, to conclude, from there, that absolute velocities are undetectable, one must \emph{assume} that only Galilean invariant quantities are detectable. The problem is that such an assumption is exactly what one wanted to prove in the first place! Hence, the standard argument is circular.

In response, one might try to argue that it is not necessary to make the assumption that only relative quantities are detectable, because such an assumption follows from the formal structure of a Newtonian universe. In section \ref{Rob} we show why this is not the case. However, we first need to say a few things regarding the notion of detection.

\section{General remarks about detections}
\label{Det}

In extremely general terms, it is said that device M detects or measures\footnote{In this work, we use the concepts of \emph{detection} and \emph{measurement} interchangeably.}  some features of system S when, in virtue of a mutual interaction, certain properties of S wind up correlated with certain properties of M.\footnote{Under this general definition---which depends strongly on the dynamics of the relevant theory---detections are not accidental events but rather modally robust correlations grounded in laws, as in \cite{Roberts}, and in counterfactuals, as in \cite{murgueitio21}.} In that case, the properties of M participating in the correlation are said to record the result of the detection. For example, when an ammeter measures the electric current in a circuit, the relative position between the needle and the dial is the objective, physical property of the ammeter that encodes the result. Similarly, when somebody sees a black cat crossing the street, such a perception gets encoded in the objective, physical state of their brain. Of course, for a complete analysis of measurements, many more things, regarding, e.g., faithfulness, observability, control or systematicity, must be said; but these fundamental features of measurements are enough for us to show the inadequacy of the standard arguments for the undetectability of absolute velocity.

The description of a detection presented above makes it clear that the \emph{dynamics} of a theory is what determines which devices can measure what properties: for some property of S to be detectable by M, the appropriate interaction between S and M must be available. Moreover, it is the dynamics that determines what degrees of freedom of M are the ones that are able to codify the result of a detection; and the important point we want to make is that, in principle, all objective, physical properties of a system could participate in the encoding of a measurement. That is, unless there is a formal restriction within a theory or an external stipulation to the contrary, all objective, physical properties could, in principle, be involved in the representation of a result.

With this in mind, consider a world governed by some theory with a given set of degrees of freedom, and suppose that those degrees of freedom can naturally be divided into different types (e.g., position and velocity, absolute and relational, field and particle, material and gravitational, field A, field B and field C, etc.). Suppose, moreover, that one would like to \emph{stipulate} that the results of all measurements can be recorded in terms of only a given type of degrees of freedom (e.g., only in terms of positions, particles, relational features, etc.). In order for such stipulation to be \emph{justified}, one must make sure that the dynamics of the governing theory is such that it allows for the pertinent correlations to be established.

Suppose, then, that system S interacts with device M. If one wants to assert that the result of any measurement of M over S can be recorded in terms of, say, the $r$-type degrees of freedom of M, then one must be able to show that an interaction between M and S can be established, such that it results in a correlation between any objective property of S and the $r$-type degrees of freedom of M. If that is not the case---i.e., if there is some objective property of S for which the dynamics does not allow for such a correlation to be established---then, one would not be able to conclude that such an objective property of S is undetectable; instead, the conclusion would have to be that \emph{to assume} that the result of all measurements can be recorded it terms of $r$-type degrees of freedom is simply \emph{unjustified}.

To see this, consider a world populated by particles with two types of properties, A and B, such that there are no inter-type influences: A-properties only interact with A-properties and B-properties only with B-properties. Suppose, moreover, that one imposes the assumption that the results of all measurements must be recorded in terms of, say, A properties. Of course, due to the decoupling between A and B properties, B properties would seem undetectable. It is clear, though, that this inability to detect B properties does not arise due to some fundamental unobservability of such properties, but from the arbitrary stipulation that all results must be recorded in terms of A properties. In fact, by construction, both A and B properties are equally real and objective, so there is no reason whatsoever to treat them differently regarding detection, nor to favor one over the other.
	
For the case of the Newtonian universe under consideration, it is not the case that all relational properties completely decouple from all absolute properties. It is only the absolute velocity of the center of mass of a system that fully decouples from all relational properties. However, given that the absolute velocities of the individual components of the system are a function of the absolute velocity of the center of mass of the full system, it is clear that absolute velocities, in general, will not be recordable in terms of relative properties.
	
To see this in more detail, let us consider the detection of absolute velocity in a Newtonian universe, in which all forces only depend on relative quantities. We consider, then, a particular system, M, and ask whether it could detect the absolute velocity of a second object, S, with which it interacts. To do this, we consider two scenarios related by an active Galilean boost. For concreteness, in scenario 1, the absolute velocity of M, after the interaction with the object, is zero. Of course, in scenario 2, M has some non-zero velocity after it interacts. Now, given that, by stipulation, all forces depend on relative quantities, it is certainly the case that all relational properties of M are identical in both scenarios. Therefore, if the result of the detection performed by M is fully encoded in purely relational properties of M, then M would necessarily record exactly the same result in both scenarios; given that the absolute velocity of the object is different in scenarios 1 and 2, under such circumstances, M would be unable to measure such an absolute velocity. However, why would it be the case that the result of the detection performed by M must be fully encoded in purely relational properties of M?
	
As we explained above, in general, such an encodement is performed in terms of some subset of the objective, physical properties of the apparatus and, in our example, there is an objective, physical property of M which differs between scenarios 1 and 2, namely the absolute velocity of its center of mass. Moreover, there does not seem to be any fundamental reason, from the internal standpoint of the Newtonian universe in question, for which such an objective, physical property could not be included in the subset of properties encoding some measurement for some apparatus. Of course, it could simply be stipulated ``by hand'' that absolute properties are never involved in detections, but that would amount to impose, not prove, that absolute velocities cannot be measured.

\section{A couple of influential analyses}
\label{Rob}
	
\citet{Roberts} contains a thorough discussion of this issue and offers an influential argument in favor of the claim that, in a Newtonian universe, absolute velocities are undetectable. In developing the argument, it is readily acknowledged that, in order to prove such an assertion, one must \emph{assume} that, for all measurements, ``results are encoded in physical quantities that are themselves invariant under the dynamical symmetries''. That is, it must be assumed that all pointer variables necessarily are Galilean invariant. However, \cite{Roberts} does not take the inclusion of such an assumption as an act of question-begging, as it deems the assumption to follow from an important fundamental feature of our world---a feature which physicists are entitled to take for granted. 
	
The idea is to try to construct an argument to show that the assumption must be true; i.e., to show that all perceptible facts must be preserved under the dynamical symmetries of our world. To do so, Roberts starts by noting that it does seem to be the case that we cannot perceive facts that are not preserved by the dynamical symmetries. Moreover, he points out that, in order for a detection to count as an empirical measurement, the results must be communicable in a public medium. From this, he designs a scenario to argue that, if we were to perceive variant facts, then things would be quite strange.

The argument is the following. Suppose that, in world $U_1$, Sally is able to accurately detect her absolute velocity and communicate the result by sending to Harry a letter (or some other message composed of purely invariant quantities). Consider now world $U_2$, constructed by applying a Galilean boost to $U_1$. Given that Sally's velocity is affected by the boost, but
the message is not, Sally's report would not be accurate in $U_2$. Maybe, Sally could send a message employing another medium containing variant properties. However, even if that could be the case, many of the standard means of communication (e.g., letters) would not be available to reliably report a certain type of information (e.g., absolute velocity). It follows that, either there are things that Sally can perceive, but not communicate, or that she cannot use certain standard media to communicate all information. However, Roberts alleges, we are not like that: creatures in those circumstances would have communicative abilities very different from our own. From that, it is concluded that the assumption required to prove that absolute velocity cannot be measured follows from very general and fundamental truths of our `form of life'. 

We find the above line of reasoning problematic. To begin with, as we mentioned earlier, one must not confuse the empirical fact that our universe seems to respect the principle of relativity with the question of whether absolute velocities are detectable in a Newtonian universe. That is, one should not employ empirical features of our world, which is patently not Newtonian, to try to determine what properties would be detectable in such a universe. As a result, particular features about how communication among humans works, are simply the wrong sort of data to explore the issue at hand.

On top of this, the ``strange'' feature uncovered through Sally's scenario is not really as mysterious as it might seem and, more importantly, has absolutely nothing to do with certain properties being detectable or not. The basic characteristic behind Sally's troubles communicating her result is the fact that, in a Newtonian universe with forces only depending on relative quantities, the center of mass of the universe fully decouples from all relative properties. But from this dynamical fact one simply cannot conclude that absolute properties are undetectable.

To see why this is so, recall the example of particles with non-interacting properties A and B described above. In that case, a system of particles would have A-properties, B-properties and even relational AB-properties. Suppose, now, that such a system measures or detects something. If so, the result would end up encoded as a function of such properties; and it seems clear that the system would be able to record results of, both, A-properties and B-properties. However, it is also clear that, due to the lack of inter-type interactions, it would be impossible for, say, an measurement of an A-property to be encoded in terms of B-properties. That is, one can easily develop an argument, analogous to Sally's, in which two worlds with different A-properties, but identical B-properties, would demonstrate the inaccuracy of this sort of measurement in most worlds. 

In fact, what is really going on is not that the measurements would be inaccurate but that, due to the decoupling of properties, it would simply be impossible to set up a procedure to record one type of property in terms of the other. This is just a natural consequence of the decoupling. The important point, though, is that this inability to record one sort of property in terms of the other has absolutely nothing to do with the issue of what properties are measurable and which are not. As argued above, A-properties and B-properties are both equally real and objective, so there is no justification for treating them differently in terms of what can be detected or for giving preference to one over the other.

Going back to the Newtonian case, the fact that the center of mass decouples from relative degrees of freedom in such a universe does mean that the result of a measurement of absolute velocity cannot end up recorded in terms of relative features. However, it does not imply that only relative properties are detectable; both absolute and relative properties are real and objective, and there is simply no way to conclude, out of the formal features of a Newtonian universe, that absolute and relative properties would behave differently regarding detection. The upshot of all this is that there are no formal reasons to conclude that absolute velocities are undetectable in such a Newtonian universe: standard arguments in favor of such an assertion are question-begging.

Another influential analysis of the subject is found in \cite{wallace}, which offers a more formal treatment of the general relation between dynamical symmetries and (among other things) observability. A central conclusion of that work is that epistemic claims about the observability of variant quantities can, in fact, be derived from purely dynamical considerations regarding symmetries.

Despite this general conclusion, Wallace seems to agree with our assessment of the analysis in \cite{Roberts}, noting that:
\begin{quote}
	...dynamical reasons prohibit us from encoding variant features of a system in its invariant features. But this generally has not been taken to resolve the question of unobservability, unless we make a supposedly question-begging assumption---after all, variant features can be encoded in other variant features. (Roberts, for instance, then appeals to anthropic features of the sort of beings we are to complete the story.). \citep[section 13.5, p. 335]{wallace}
\end{quote}
However, by imposing what we take to be an unjustified restriction on the class of admissible measurement records, he ultimately reaches the opposite conclusion from ours. Let us unpack this in more detail.

In \cite{wallace}, the discussion is formalized by considering the action $R$ of the symmetry $G$ and defining the \emph{orbit} $O$ of this action as the equivalence classes under the relation: $x \sim x'$ if and only if $x'= R(g)x$ for some $g \in G$. Then, by picking an arbitrary reference point $\phi_O$ in each orbit, any $x \in O$ can be expressed as $x = R(g)\phi_O$ for some $g$, and hence, by the ordered pair $(O, g)$. The value of all this is that the action of the symmetry can be expressed as
\begin{equation}
R(h)(O,g)=(O,hg),
\end{equation}
which means that it decomposes into its $G$-invariant part, represented by the orbit $O$, and its $G$-variant part, represented by an element of $G$ itself.

With this framework in place, Wallace first considers the case of an isolated system possessing a dynamical symmetry and asks whether variant properties could be observable from within such a system. At the crucial step, however, he simply stipulates---without adequate justification, dynamical or otherwise---that a `readout' quantity associated with a detection, ``must be independent of $g$'' \cite[p. 329]{wallace}. From this stipulation, his desired conclusion follows immediately: ``no measurement internal to a system can distinguish whether a symmetry has been applied to the whole system,'' and hence ``the Unobservability [of variant quantities] is a straightforward consequence of the dynamics'' \cite[p. 330]{wallace}.

Next, by introducing a detection model practically identical to ours, Wallace extends the analysis in several directions, including a composite system consisting of a target system and a measuring device, where $G$ is also a dynamical symmetry of the measuring device---and hence of the composite system as a whole. He then observes that, in this setting, the possibility of measuring a variant quantity ultimately reduces to the possibility of measuring a $G$-variant quantity of the composite system from within. Finally, appealing to the previous result for isolated systems, he concludes that, also in this composite scenario, variant quantities are undetectable. Note, though, that since the former result rests on an unwarranted stipulation, the latter inherits the same weakness.

In sum, the conclusion reached in \cite{wallace}---namely, that variant quantities are undetectable---can be traced to the stipulation that detection readouts must be independent of variant quantities. However, this restriction is introduced without adequate justification and, as we argued above, no such restriction can be derived from dynamical or purely formal considerations. We therefore conclude that the conclusion reached in \cite{wallace} is unwarranted.

\section{Correlations, detections and observations}
\label{Mis}

The discussion above might incorrectly reinforce the impression that, via purely relational interactions, it would be impossible to establish correlations between absolute properties---and, thus, that it would be impossible to detect them. It is true that the center of mass of a system completely decouples from its relational properties and, as a consequence, absolute velocity cannot be recorded in terms of relational quantities. However, this does not imply that, via purely relational interactions, one cannot setup a scenario in which the \emph{absolute} properties of two systems end up correlated after they interact. That is, even if all interactions are fully relational, the final absolute properties of M, some measuring device, can encode---i.e., be correlated with---the initial absolute properties of some system S. 

To see this, consider a set of runs in which a measuring device M, with fixed initial conditions, interacts with system S, with varying initial conditions (suppose, for simplicity, that M and S are isolated from the rest of the world). Now, given that the interaction between M and S only depends on relational quantities, it is true that the absolute velocity of the center of mass of the M+S system will fully decouple from all relational properties between them---and thus not change during the interaction. However, that does not mean that the individual absolute velocities of the centers of mass of M and S will remain constant as well. In fact, even though the interaction between M and S is completely relational, the final absolute velocity of M will, in general, be a function of the initial absolute velocity of S. In other words, the final absolute velocity of M will absolutely be able to encode---and thus measure---the initial absolute velocity of S.\footnote{Although \cite{Dasgupta2016} agrees with us that absolute velocities can get correlated in this way, it still endorses Roberts' argument for the undetectability of absolute velocities.}

The point is that, while the fact that all interactions are relational does imply that one cannot record the result of a measurement of an absolute property in terms of a relational one, it does not mean that absolute velocities themselves cannot get correlated during a purely relational interaction. Therefore, through such a sort of process, it is perfectly possible to record---i.e., measure---the initial absolute velocity of a system in terms of the final absolute velocity of a measuring device.

For a very simple illustration, consider a one-dimensional elastic collision between M and S (with {\emph{known} masses $m_M$ and $m_S$). In that case, $v_M^f$, the final absolute velocity of M, satisfies
\begin{equation}
v_M^f = \frac{m_M-m_S}{m_M+m_S} v_M^i+ \frac{2m_S}{m_M+m_S} v_S^i
\end{equation}
with $v_M^i$ and $v_S^i$ the initial absolute velocities of M and S. Then, for a \emph{known}, \emph{fixed} $v_M^i$, $v_M^f$ conveys full information of $v_S^i$---i.e., M measures the initial velocity of S, encoding it in $v_M^f$. Moreover, if an active Galilean boost (with velocity $v$) is performed on the whole M+S system
\begin{equation}
v_M^i \rightarrow v_M^i + v , \qquad v_S^i \rightarrow v_S^i + v,
\end{equation}
then $v_M^f$, the quantity encoding the result, transform exactly as it should
\begin{equation}
v_M^f \rightarrow v_M^f + v.
\end{equation}

At this point, it might be argued that all we have accomplished is to show that adequate correlations between absolute properties can be established, but that correlations are not sufficient for detection. This, in fact, only contributes to our cause. What we have shown is that, within a Newtonian universe, relative and absolute velocities behave in exactly the same way regarding all relevant features: in both cases, adequate measurement protocols can be performed, the necessary correlations can be established, all the required counterfactuals are satisfied, etc. It follows that there is no principled way, within the Newtonian universe, to conclude that absolute velocities are undetectable. One might still insist that relative quantities possess something extra, which makes the difference. Our point is that, if it exists, this extra ingredient is not to be found within the Newtonian framework. That is, (possibly motivated by a search for empirical adequacy,) one could stipulate that, even though relative and absolute velocities have the exact same set of relevant features, only relational quantities are detectable; but that would be no more than a stipulation, with no internal justification.

We do not claim to have shown that absolute velocity is detectable; we have shown, instead, that formal arguments to the effect that absolute velocity is undetectable beg the question. This contrasts with the standard claim that the undectability of absolute velocity follows from the mere formal structure of Newtonian dynamics. Of course, there are many differences between absolute and relative quantities, but the point is that, contrary to what is usually asserted, none of those differences can be leveraged to construct an argument for the undetectability of absolute velocity.

\section{Broader consequences}
\label{Cons}

By now, it is quite clear that the world is not Newtonian. Therefore, one may doubt the practical value of exploring whether absolute velocity would be detectable in a Newtonian universe. While such a question might seem too academic and narrow, the analysis presented above contains an important, general lesson: one must always treat with suspicion any claim within a physical theory, regarding the existence of real-but-undetectable entities. The problem is that the arguments for such a claim  probably contain the same shortcomings uncovered above with respect to arguments for the undetectability of absolute velocity. That is, if an entity is deemed real and objective within a physical theory, then any claim regarding its undetectability in all likelihood depends upon an arbitrary stipulation as to what can be detected or observed.

In fact, in \cite{ExposingPilotwave2025}, using an analogous analysis to the one developed here, it is shown that the alleged limited empirical access to the positions of the particles within the de Broglie-Bohm pilot-wave theory \citep{AU}, depends upon the debatable assumption that the results of all measurements are always configurationally grounded, i.e., that all relevant information about a system can always be encoded in the positions of the particles in its environment. In particular, it is argued that, similar to the Newtonian case explored here, the motivation behind the imposition of a purely configurational grounding of information relies on circular reasoning or unjustified appeals to empirical evidence. 
	
Beyond physics, our analysis seems to cast doubt regarding any alleged case of true, in principle, indiscernibility between different physical theories. If so, it may have interesting consequences for well-known debates in general philosophy of science. Thus, what might appear to be a narrow analysis in the context of an old physical theory, brings with it far-reaching implications for both physics and general philosophy of science. 

\section{Conclusions}
\label{Conc}

We have shown that standard arguments, based solely on the formal features of a Newtonian universe, which claim that absolute velocity would be undetectable in such a universe, beg the question. Such arguments, instead of deriving the undetectability of absolute properties out of the formal principles of Newtonian mechanics, (implicitly or otherwise) simply \emph{assume} the undetectability of absolute properties. It is true that, if all forces are assumed to only depend on relational properties, then a measurement of an absolute property cannot be recorded in terms of relative properties. However, from this does not follow that absolute properties cannot be detected. In fact, we have show that, via purely relational interactions, the initial absolute velocity of a system S can be reliably recorded in the final absolute velocity of a measuring device M. 

To conclude, we stress the point not to confuse a purely theoretical question regarding what can be shown to be undetectable from the formal principles of a given theory in a hypothetical scenario with empirical facts regarding what can actually be detected in our universe. In particular, actual empirical facts should never contaminate the theoretical analysis, nor lead to motivated reasoning in order to unwittingly impose empirical equivalency between some framework and the world. That is, if the results of an analysis of what can be detected within some framework do not conform to what we actually are able to detect, then the lesson to be learned is not that a restriction on the type of records allowed can be imposed arbitrarily, but that the model in question is not empirically adequate.



\bibliographystyle{apalike} 
\bibliography{bibNew.bib}
\end{document}